\documentclass[preprint,aps,showpacs]{revtex4}
\usepackage{epsfig}

\begin{document}

\draft

\title{Neutron transfer reactions induced by $^8$Li on $^9$Be}

\author{V. Guimar\~aes, R. Lichtenth\"aler, O. Camargo, A. Barioni,}
\affiliation{Instituto de F\'{\i}sica, Universidade de S\~ao Paulo 
P.O.Box 66318, 05389-970 S\~ao Paulo, SP, Brazil.}

\author{M. Assun\c c\~ao,}
\affiliation{Universidade Federal de Sao Paulo, Campus Diadema,
09941-510 S\~ao Paulo, SP, Brazil.}

\author{J. J. Kolata,}
\affiliation{Department of Physics, University of Notre Dame, Notre Dame,
Indiana, 46556, USA.}

\author{H. Amro, F. D. Becchetti, Hao Jiang,}
\affiliation{Department of Physics, University of Michigan, Ann Arbor, 
Michigan 48109-1120, USA.}

\author{E. F. Aguilera, D. Lizcano, E. Martines-Quiroz, and H. Garcia.}
\affiliation{Instituto Nacional de Investigaciones Nucleares, A. P. 18-1027, 
C.P. 11801, Distrito Federal, Mexico.}

\date{\today}

\begin{abstract}
Angular distributions for the elastic scattering of $^8$Li on $^9$Be
and the neutron transfer reactions $^9$Be($^8$Li,$^7$Li)$^{10}$Be and 
$^9$Be($^8$Li,$^9$Li)$^8$Be 
have been measured with a 27 MeV $^8$Li radioactive
nuclear beam. Spectroscopic factors for $^8$Li$\otimes$n=$^9$Li and 
$^7$Li$\otimes$n=$^8$Li bound systems
were obtained from the comparison 
between the experimental differential cross section and finite-range 
DWBA calculations with the code FRESCO.  
The spectroscopic factors obtained are compared to
shell model calculations and  to
other experimental values from (d,p) reactions. Using the present
values for the spectroscopic factor, cross sections 
for the direct neutron-capture reactions $^7$Li(n,$\gamma$)$^8$Li and
$^8$Li(n,$\gamma$)$^9$Li were calculated in the framework
of a potential model.
\end{abstract}

\pacs{21.10.Jx, 25.60.Je, 25.40.Lw}

\maketitle


\section{Introduction}

Spectroscopic study of nuclei far from the valley of $\beta$ stability
is one of the major active fields in nuclear physics.
To perform nuclear spectroscopic investigation on such nuclei,
direct reaction processes that add or remove one or a few nucleons, 
such as direct stripping  and pickup reactions, from which one can 
identify single-particle orbitals, energies and their occupancies are 
commonly used.
With the improvement of radioactive ion beam intensities it is now possible 
to obtain reliable measurements of transfer cross sections induced by 
unstable projectiles, and  together with finite-range computer codes 
such as FRESCO \cite{thom88},  it becomes possible to obtain spectroscopic 
information on unstable nuclei with good precision.

In particular, spectroscopic investigation of unstable lithium isotopes 
is of interest not only for understanding nuclear structure and reaction 
mechanisms, where 
spectroscopic factors of these isotopes can be an important experimental 
probe for  single-particle dynamics, shell model calculations,
and halo properties \cite{nak06,mich05,tsang05,nav04,santi03}, but
also for astrophysics. 
In the  inhomogeneous nucleosynthesis models \cite{kaj90}, reactions 
with $^8$Li  can bridge the A=8 mass gap and heavier elements would then be 
synthesized in the early universe. 
It also has been pointed out that neutron-induced  three-particle interactions 
can be important  in a high neutron abundance environment \cite{gor95}. 
In this case, heavier elements would 
be synthesized via the reaction sequence  
$^4$He(2n,$\gamma$)$^6$He(2n,$\gamma$)$^8$He($\beta^+$)$^8$Li.
Here again, $^8$Li plays an important role in the subsequent synthesis of 
heavier elements. 
Likewise, light neutron-rich nuclei such 
as $^8$Li have been found to be important to produce seed nuclei  for the
r-process, e.g., in Type-II supernovae \cite{ter01}.

Once $^8$Li is  produced two possible competing chains of reactions
can take place:
$^8$Li($\alpha$,n)$^{11}$B(n,$\gamma$)$^{12}$B($\beta^+$)$^{12}$C  and 
$^8$Li(n,$\gamma$)$^9$Li($\alpha$,n)$^{12}$B($\beta^+$)$^{12}$C.
The key reaction for the second chain is $^8$Li(n,$\gamma$)$^9$Li. 
Direct measurement of this  reaction is impossible  because no neutron  
target exists and the half-life of  $^8$Li is too short (838 ms) for $^8$Li 
to be used  as a target.   
Thus, experimental information on the cross section
for this reaction has to be obtained by some indirect methods. 
Kobayashi {\it et al.} used the Coulomb dissociation method \cite{kob03}, 
which is  the inverse of the capture process. However, in this work only an 
upper limit for the cross section at low energy was obtained. 
Another indirect approach that may be used to investigate capture reactions 
is the 
ANC (Asymptotic Normalization Coefficient) method, where the (d,p) reaction 
in inverse kinematics is used to extract an ``asymptotic normalization 
coefficient" that can be related to the capture cross section.
The $ANC$ is obtained from peripheral  transfer reactions whose 
amplitudes contain the same overlap function as the amplitude of the  
corresponding capture reaction of interest \cite{gag02} and therefore
can be used to normalize the non-resonant part  of the capture reaction.
The method is based on the assumption that capture reactions at stellar 
energies usually proceed through the tail of the nuclear overlap function. 
The amplitude of  the radiative capture cross  section is then  
dominated by contributions from large relative distances of the  
participating nuclei.  However, it
has been shown that s-wave neutron capture, even at rather low energies,
is not peripheral \cite{men95,nag05} and so it is necessary to use a 
potential model to calculate the wave function of the incoming neutron
in this case. 
This indirect aproach, based on the
potential model, has been recently applied in the analysis
 of the 
$^{16}$O(d,p)$^{17}$O and $^{16}$O(d,n)$^{17}$F transfer reactions 
to determine the corresponding $^{16}$O(p,$\gamma$)$^{17}$F
and $^{16}$O(n,$\gamma$)$^{17}$O
astrophysical direct capture cross sections \cite{ass06}.

Early measurements of neutron transfer reactions induced by stable lithium 
isotopes on a $^9$Be target ($^9$Be($^7$Li,$^6$Li)$^{10}$Be \cite{kemp77} and
 $^9$Be($^6$Li,$^7$Li)$^{8}$Be \cite{cook85}) have shown that these
reactions are good tools to obtain spectroscopic information.
In this paper, we report on the measurement and analysis of angular
distributions for two neutron-transfer reactions induced by a
radioactive $^8$Li beam: $^9$Be($^8$Li,$^9$Li)$^8$Be and  
$^9$Be($^8$Li,$^7$Li)$^{10}$Be. From a FR-DWBA analysis of these angular 
distributions, the spectroscopic factors for the 
$^7$Li$_{gs}\otimes$n=$^8$Li$_{gs}$ and 
$^8$Li$_{gs}\otimes$n=$^9$Li$_{gs}$ 
bound systems
were extracted, and cross sections
and nucleosynthesis reaction rates for 
the non-resonant part of  the $^7$Li(n,$\gamma$)$^8$Li 
and $^8$Li(n,$\gamma$)$^9$Li capture reactions were derived. 

\section{The experiment }

Angular distributions for $^8$Li elastic
scattering and the $^9$Be($^8$Li,$^9$Li)$^8$Be and 
$^9$Be($^8$Li,$^7$Li)$^{10}$Be
neutron transfer reactions have been measured at the Nuclear 
Structure Laboratory of the University of Notre Dame, USA.
The 27 MeV secondary $^8$Li radioactive beam was obtained from the 
{\it TwinSol} RNB system \cite{bec03}. In this system, the 
beam is produced in a primary target via the $^9$Be($^7$Li,$^8$Li) reaction, 
where a 30 MeV primary $^7$Li beam having an intensity of up to 1 e$\mu$A
was obtained from a 9.5 MV Tandem Van de Graaff Accelerator. 
The production target consisted of a beryllium foil 12 $\mu$m thick mounted 
as the exit window of a gas cell which was filled with $^4$He gas at a 
pressure of 1 atm for cooling purposes (the entrance window of the gas cell 
consisted of a 2 $\mu$m thick Havar foil).  
The two superconducting  solenoids in the {\it TwinSol} system act as 
thick lenses to collect, select, and focus the secondary beam into a 
scattering chamber.  The 27 MeV $^8$Li beam had an average intensity of  
5.0 $\times$ 10$^5$ particles per second per 1 e$\mu$A of primary beam,
and an energy resolution of 0.450 MeV (FWHM) determined from the elastic 
scattering measurement on a  gold target. The beam was focused onto a  
1.44 mg/cm$^2$ thick $^9$Be secondary target. Some beam contaminants 
($^4$He, $^6$He and $^7$Li) with the same magnetic rigidity as $^8$Li
were also present
but did not produce reaction products with mass A=8 or A=9 in the same range 
of energy as the particles from the neutron transfer reactions of interest.

The scattered $^8$Li particles and $^7$Li and $^9$Li reaction products were
detected by an array of $\Delta$E-E Si telescopes. 
The measurements were performed with two setups, where a combination 
of the 3 telescopes covered the range of laboratory angles 
from 15-40 degrees.  An overlap of angles in these setups was useful 
for normalization purposes. The telescopes consisted of 20-25 $\mu$m Si 
$\Delta$E detectors backed by 300 $\mu$m thick Si E detectors. 
The detector telescopes  had circular apertures which subtended a solid 
angle of 4 msr for the most forward-angle measurements and 8-15 msr 
for the  backward angles. 
A collimator placed in front of the  first solenoid limited the angular 
acceptance of the particles produced in the primary target to the range 
of $2.5^0-6.0^0$. This introduces an angular divergence in the secondary beam 
of about $\pm 3^0$.
Since the angular aperture of the collimators in front of the 
detectors was also about $\pm 3^0$, the average detection angle were determined
by a Monte Carlo simulation, which took into account the collimator
size in front of the detectors, the secondary beam spot size on the 
secondary target (4mm), the 
secondary beam divergence and the angular distribution in the range of 
the  detector aperture (Rutherford on gold and calculated in an iterative way
for the $^9$Be target).
 
The simultaneous measurement of the transfer 
products and  elastic scattering was very useful to check the consistency 
of the  overall normalization and to select an optimal set of optical-model 
potential parameters. The latter are very important in the FR-DWBA transfer
calculations. During the experiment $^8$Li elastic scattering on a 
gold target, which was pure Rutherford, was also measured to obtain the 
absolute normalization of the data.

\section{Data Analysis} 

Reaction products were identified 
using  a two dimensional [$C(Z,M) \times E_{total}$] plot. 
The particle identification constant, $C(Z,M)$, is given by:
$C(Z,M)=(E_{total})^b - (E_{total}-\Delta E)^b$ \cite{knoll}, where
$E_{total}=\Delta E + E_{residual}$ and  $b=1.70$ for these 
light particles.
This constant appears as straight lines as a function of the energy for 
each Z and M particle in the two dimensional spectrum
 [$C(Z,M) \times E_{total}$]. 
A typical  particle identification spectrum 
is shown in Fig. 1 for the lithium and helium isotope region. 
In this plot the $^8$Li scattered beam particles and the $^9$Li and $^7$Li 
reaction products are shown and could be easily identified.

\subsection{Elastic scattering}

The $^8$Li energy spectra were obtained by selecting and projecting 
the $^8$Li region in the [$C(M,Z) \times E_{total}$] plot.
The experimental resolution, 0.450 MeV (FWHM),
was sufficient to separate elastic scattering from
inelastic scattering to the first excited state of $^8$Li (E$_x$=0.980 MeV).
The experimental angular distribution obtained for the elastic scattering 
of $^8$Li on  $^9$Be is shown in 
Fig. 2. The data were analyzed with the optical model (OM) using 
volume-type Woods-Saxon nuclear potentials and Coulomb potentials due 
to uniform charged spheres. The optical parameters used to describe 
the  $^9$Be($^8$Li,$^8$Li) elastic scattering  are from Refs. \cite{kemp77} 
and \cite{perey76} and they are listed in Table I. The results of the 
OM calculations using these potential parameters 
can be seen in  Fig. 2. 
The OM calculations show more oscillations than the 
elastic scattering data. In the Fig. 2-b we show the calculated angular
distributions smeared  by the range of the angular
aperture and beam angular dispersion ($\Delta \Theta$=10 deg. in the C.M.).
As one can see, the comparison with the data is improved by 
smearing the calculated angular distributions. 
The curve indicated as the SP-Potential
corresponds to OM calculations using the Sao Paulo 
Potential \cite{cha02},  which is a double-folding potential with energy 
dependence and  non-locality correction.  Including a spin-orbit term,  
$V_{SO}=7.0$ MeV, in the optical potential did not change the results. 
Although there was no attempt to adjust the parameters to 
fit the data, the calculations with all of these potentials give a good 
description of the elastic scattering data.  The SP-Potential also reproduces
quite well the absolute normalization, which is of some interest considering 
that this folding-model potential has no free parameters.

These potentials 
were used in the DWBA calculations for the transfer reactions as
described in the following sections.

\subsection{Neutron stripping reaction: $^9$Be($^8$Li,$^7$Li)$^{10}$Be}

The  $^9$Be($^8$Li,$^7$Li$_{gs}$)$^{10}$Be$_{gs}$ 
reaction has a positive Q-value 
of +4.780 MeV. Thus, the $^7$Li particles from this transfer 
reaction could easily be separated from the elastic $^8$Li particles
in the $C(M,Z) \times$ Energy plot. 
At the most forward angles (15 and 18 degrees) the $^7$Li group was
double peaked due to inelastic scattering to the first excited 
state of $^7$Li$^{*}$ at Ex=0.470 MeV.
Although the overall experimental energy resolution (0.450 MeV) was 
barely sufficient to resolve the $^7$Li ground state from the first 
excited state, it appears that ($^8$Li,$^7$Li$_{gs}$) was the 
dominant neutron-transfer mode.  At backward angles,
the $^7$Li groups observed were centered at the g.s. energies and 
were not double peaked, at least within the statistics obtained. 
At the most forward angles, the second excited state of $^7$Li 
(E$_x$=2.70~MeV) was also observed but is not considered in the present 
analysis. The angular distribution for the 
$^9$Be($^8$Li,$^7$Li$_{gs}$)$^{10}$Be$_{gs}$ reaction 
is shown in Fig. 3. As can be seen, the differential cross sections for 
this transfer process are  not very large (less than
1 mb/sr) which made the measurements quite difficult at the backward 
angles due to the limited secondary beam intensity.

Finite-range distorted-wave Born approximation (FR-DWBA) 
calculations for the $^9$Be($^8$Li,$^7$Li)  neutron transfer reaction have 
been performed using the  code FRESCO \cite{thom88}. 
For each FR-DWBA calculation, the same optical-model potential 
parameters (Table-I) were used for both entrance ($^8$Li+$^9$Be) and 
exit ($^7$Li+$^{10}$Be) channels.  The bound-state wave
functions were generated with Woods-Saxon potentials and geometric 
parameters $r=1.25$ fm and  $a=0.65$ fm, with the depths of the potentials  
adjusted to give the correct separation energies. 
In the present ($^8$Li,$^7$Li) DWBA calculation the neutron is considered 
to be transfered from the  $p_{3/2}$ orbital in $^8$Li$_{gs}$($J^{\pi}=2^+$) 
to the $p_{3/2}$ orbital in $^{10}$Be$_{gs}$($J^{\pi}=0^+$), leaving
$^7$Li in its ground-state $J^{\pi}={3 \over 2}^-$. There is a possibility 
of a contribution from the $p_{1/2}$ admixture in the $^8$Li ground-state 
but according to a spectroscopic factor calculation by Cohen and Kurath 
this contribution is about 5\% and it was not considered in the transfer 
reaction DWBA calculation.
One-nucleon transfer 
reactions provide a spectroscopic factor for one vertex if that of the 
other vertex is known. In the FR-DWBA calculation for the 
$^9$Be($^8$Li,$^7$Li)$^{10}$Be neutron transfer reaction, the spectroscopic 
factor for the $^9$Be$_{gs}\otimes$n=$^{10}$Be$_{gs}$ system was
taken to be 
$S_{^{10}Be}$=2.23$\pm$0.13, which is the average of values from two (d,p) 
studies \cite{pow70,dar76} (see Table-II).  By normalizing the calculation to 
the experimental data,
a spectroscopic factor  of $S_{^8Li}$(g.s.)=1.03 $\pm$ 0.15
was obtained for the $^7$Li$_{gs}\otimes$n=$^8$Li$_{gs}$ vertex. 
This value is compatible with the 
Cohen and Kurath shell model calculations \cite{coh67} (see Table-II). 
A similar situation has been found in the analysis of the 
$^9$Be($^7$Li,$^6$Li)$^{10}$Be  \cite{kemp77} and 
$^9$Be($^6$Li,$^7$Li)$^{8}$Be reactions \cite{cook85}, where the 
spectroscopic factor for both vertices involved in the transfer 
agreed with values calculated by Cohen and Kurath.   These results indicate 
that Cohen-Kurath wave functions describe stable lithium 
and beryllium isotopes in the mass range A=6 to A=10 reasonably well.

\subsection{Neutron pick-up: $^9$Be($^8$Li,$^9$Li)$^{8}$Be}

The $^9$Be($^8$Li,$^9$Li)$^8$Be neutron-transfer reaction can proceed by 
two possible contributions that leave $^9$Li in its
ground state. These correspond to neutron transfer 
to either a $1p_{1/2}$ or $1p_{3/2}$ orbit in $^9$Li. Here only transfer to 
the $1p_{3/2}$ orbit is considered because the contribution of
the $1p_{1/2}$ orbit has been found to be less 
than 5\% \cite{mohr03}.  
Finite-range distorted-wave Born approximation (FR-DWBA) 
calculations for the $^9$Be($^8$Li,$^9$Li)  neutron transfer reaction have 
been performed using also the  code FRESCO \cite{thom88}. 
For each FR-DWBA calculation, the same optical-model potential 
parameters (Table-I) were used for both entrance ($^8$Li+$^9$Be) and 
exit ($^9$Li+$^{8}$Be) channels.  The bound-state wave
functions were generated with Woods-Saxon potentials and geometric 
parameters $r=1.25$ fm and  $a=0.65$ fm, with the depths of the potentials  
adjusted to give the correct separation energies.
To obtain the spectroscopic factor for the 
$^8$Li$_{gs}\otimes$n=$^9$Li$_{gs}$ vertex, the spectroscopic factor for
$^8$Be$_{gs}\otimes$n=$^9$Be$_{gs}$ must be known.
A value of $S_{^9Be}=0.44(7)$ was used for this
vertex, which is the average
of spectroscopic factors from two (d,t) reactions \cite{dar76}, \cite{fit67}.
This is somewhat smaller than the Cohen-Kurath prediction (see Table-II).
Normalizing the FR-DWBA 
calculation to the experimental data, a spectroscopic factor of 
$S_{^9Li}=0.62 \pm 0.13$ was obtained for the  
$^8$Li$_{gs}\otimes$n=$^9$Li$_{gs}$  
bound system.  The results of the FR-DWBA calculations with different 
sets of parameters are shown in  Fig. 4.  As one can see, the calculations
agree extremely well with the data at forward angles.  The uncertainty in the 
spectroscopic factor is estimated to be about 20\% due to 
the uncertainties in the experimental data at forward angles and in
the spectroscopic factor for the $^8$Be$_{gs}\otimes$n=$^{9}$Be$_{gs}$ 
vertex.

The spectroscopic factor obtained for the  
$^8$Li$_{gs}$$\otimes$n=$^9$Li$_{gs}$
system in the present analysis is compared
with other experimental values and shell model calculations in Table-II.
Our result agrees very  well with the value from the shell model 
calculation of Ref. \cite{nav04}
and with the experimental value obtained from a (d,p) reaction at lower energy 
\cite{li05}, but is lower than the values obtained from a recent calculation
using Cohen and Kurath wave functions \cite{jep06} and from a (d,p)
reaction study at higher energy \cite{wuo05}.
An  analysis with the SP folding-model potential was also performed and 
spectroscopic factors $S_{^8Li}$=0.68 and $S_{^9Li}$=0.40 were obtained.
These values are about 35\% lower than those obtained with
SET 1,2, and 3 for both the ($^8$Li,$^7$Li) and ($^8$Li,$^9$Li) transfer 
reactions.  This may be related 
to non-locality effects which are taken into account in the folding-model 
potential
but not in the optical model potentials. Further investigation of this 
subject appears to be warranted.

To extract the spectroscopic factors from the present 
transfer angular distributions, only the angular range 
$\theta_{\rm CM}$ $\leq$ 45 degrees was considered. To verify the 
peripherality of this transfer reaction, 
the influence of
the internal part of the overlap function  was tested in the DWBA 
calculation.  The result of such a test can be seen in Fig. 5. 
Increasing the radius cut in the radial integral up to $R_{cut}$=4.0 fm
did not produce any change in the FR-DWBA calculation in the angular range
from 0-45 deg.  Only for $R_{cut}>$5.0 fm, which corresponds
to a radius larger than the distance of closest approach of the two interacting
nuclei (i.e., $R=1.25 \times(A_1^{1/3} + A_2^{1/3})$ fm), is the
calculation sensitive to the radius cut, as observed in the 
change of the trend in the curve for $R_{cut}$=5.0 fm as compared
with the other curves in Fig. 5.

\section{Radiative neutron capture reactions}

To calculate the $^7$Li(n,$\gamma$)$^8$Li and $^8$Li(n,$\gamma$)$^9$Li
radiative capture reaction cross sections, the computer code 
RADCAP \cite{bertu03}, based on a potential model, was used. 
In the potential-model framework, the direct neutron radiative capture 
by a nucleus $b$ going to a composite nucleus $c$ via a 
transition with E1 electric dipole character is given by:

\begin{equation}
\sigma^{E1}_{b\rightarrow c} (n,\gamma) = {16\pi \over 9\hbar} k^3_{\gamma}
|Q^{E1}_{b\rightarrow c}|^2,
\end{equation}

\noindent where $k_{\gamma}=\epsilon_{\gamma}/\hbar c$ is the
wave number corresponding to a $\gamma$-ray energy 
$\epsilon_{\gamma}$.
The term $Q^{E1}_{b\rightarrow c}$ is the E1 transition matrix 
element given by:

\begin{equation}
Q^{E1}_{b\rightarrow c} = <\psi_{scat}|O^{E1}|I_{bound}>,
\end{equation}

\noindent where $O^{E1}$ stands for the electric dipole operator
and the initial-state wave function $\psi_{scat}$ is the incoming
neutron wave function scattered by the neutron-nucleus
potential. 
Here the effective charge for the neutrons used in the eletric 
dipole operator is given by $e_{eff}=-eZ/A$, where $A$ is the mass 
of the compound nucleus. 
The wave functions necessary in the potential model are
obtained by solving the scattering and bound-state
systems, respectively, for a given potential.  Thus, the 
essential ingredients of the model are the potentials used to 
generate the wave functions  $\psi_{scat}$ and
 $I_{bound}$, and the normalization for the latter which is given
by the spectroscopic factor.

\subsection{The $^7$Li(n,$\gamma$)$^8$Li capture reaction}

The direct radiative capture (DRC) of a $s$- or $d$-wave neutron 
by $^7$Li, leaving the $^8$Li compound nucleus in either the ground 
state ($J^{\pi} = 2^+$)
or the first  excited state($J^{\pi} = 1^+$, E$_x$ = 0.980 MeV) proceeds
by an E1 transition. To calculate 
the capture cross section in the potential model, 
a Woods-Saxon form with radius and  diffuseness 
parameters $r_0=1.25$ fm and $d=0.65$ fm, respectively, was adopted for both
bound-state and scattering potentials.
The depth of the bound-state potentials, $V_0(g.s.)=46.38$ MeV and   
$V_0(1st.)=43.30$ MeV, were adjusted to reproduce the corresponding  
binding energies ($E_{g.s.}=2.033$~MeV and $E_{1st}=1.052$~MeV) of the 
$^8$Li ground state and first excited state, respectively.
The potentials used to describe the scattering of the neutron by the 
$^7$Li nucleus also had geometric parameters 
 $r_0=1.25$ fm and $d=0.65$ fm.  Well depths of $V_0=56.15$ MeV and  
$V_0=46.50$ MeV  were used for the two channel-spin components 
$s=2$ and $s=1$, respectively.
These potentials were obtained by Nagai {\it et al.} \cite{nag05} by 
adjusting the well depths to reproduce the experimental n+$^7$Li scattering 
lengths ($a_+=-3.63$ and $a_-=+0.87$) of the two channel-spin
components.  To calculate the cross sections for the
$^7$Li(n,$\gamma$)$^8$Li capture reaction, leaving $^8$Li in 
the ground state ($J^{\pi}=2^+$) and first excited state($J^{\pi}=1^+$),
we used the  same potential parameters for the scattering potentials
and  the same spectroscopic factors (i.e., $S_{^8Li}$(g.s.)=0.87 and 
$S_{^8Li}$(1st)=0.48  
for the $^7$Li$_{gs}\otimes$n=$^8$Li$_{gs}$ and 
$^7$Li$_{gs}\otimes$n=$^8$Li$_{1st}$, 
respectively) as obtained by  Nagai {\it et al.} \cite{nag05}.
The results of this calculation, assuming only an $s$-wave 
neutron capture, are compared in Fig. 6 with the experimental data from Refs. 
\cite{nag05}, \cite{hei98}, \cite{nag91}, \cite{wie89} and \cite{imh59}.
The $s$-wave direct neutron
capture contribution is the dominant process, and the contribution 
from the $d$-wave neutron is found  to be negligible (less than 0.5 percent) 
at these low energies. The upper solid curve in Fig. 6 was obtained by 
using the spectroscopic factor $S_{^8Li}$(g.s.)=1.03 (15) for the 
$^7$Li$_{gs}\otimes$n=$^8$Li$_{gs}$ bound state obtained from 
the present analysis of the $^9$Be($^8$Li,$^7$Li) transfer reaction.
The different potentials used for the two different couplings of the
entrance channel spin were crucial in the calculation to correctly 
describe the known data.
As can be seen, this produces a better agreement of the calculation 
with the low-energy data. 
Once we have verified that the procedure to obtain the parameters used in the 
potential model calculation reproduced the experimentally known cross 
section for the $^7$Li(n,$\gamma$)$^8$Li capture reaction, we now extend this 
procedure to calculate the cross section for the 
$^8$Li(n,$\gamma$)$^9$Li$_{gs}$ capture reaction.

\subsection{The $^8$Li(n,$\gamma$)$^9$Li capture reaction}

As mentioned above, the radiative neutron capture cross section for the
$^8$Li(n,$\gamma$)$^9$Li$_{gs}$ reaction cannot be obtained from a direct 
measurement. We here use the same prescription, 
based on a potential model, that was followed to calculated the cross section 
for the $^7$Li(n,$\gamma$)$^8$Li reaction. The capture cross sections were 
calculated  assuming that the $^8$Li(n,$\gamma$)$^9$Li$_{gs}$
reaction proceeds by direct $E1$ capture of an $s$-wave neutron by 
$^8$Li($J^{\pi}=2^+$) leading to the $^9$Li($J^{\pi}=3/2^-$) ground 
state.
The optical-model potential used to generate the distorted wave for neutron 
scattering on the unstable  $^8$Li nucleus cannot be obtained experimentally. 
Instead, we used the same volume integral per nucleon as that of the 
n-$^7$Li potentials.  Experimental optical-model parameters can vary strongly
for different systems, 
but the volume integral per nucleon of a potential is known to have less 
ambiguity  and thus it can be considered a more stable quantity as a function 
of the masses of the interacting nuclei \cite{sat90,bec69,li05}.  
Thus, by scaling the real depth of the two entrance-channel spin
components, $s=1$ and $s=2$, from the known n-$^7$Li potential 
in the previous analysis, and keeping the same  
volume integrals of the potential per nucleon ($J_V/A=793$ MeV fm$^3$ 
and $J_V/A=657$ MeV fm$^3$, respectively), $V_0=58.15$ MeV and 
$V_0=48.15$ are obtained for the depths of the potential wells
for the two channel-spins ($s=5/2$ and $s=3/2$, respectively) 
of the n(1/2$^+$)+$^8$Li($2^+$) system. 
To calculate the overlap integral of the $^8$Li$\otimes$n=$^9$Li$_{gs}$ 
bound-system, a Woods-Saxon shaped potential is also considered. 
A depth $V_0=47.82$ MeV was 
obtained for potential by adjusting it to reproduce 
the binding energy (E=4.064 MeV). 
The normalization of this overlap integral is giving by
the spectroscopic factor $S_{^9Li}$(g.s.)=0.62 (7) extracted from the present 
analysis of the  $^9$Be($^8$Li,$^9$Li)$^8$Be transfer reaction.
With these potentials and the spectroscopic factor, the cross sections
for the non-resonant
part of the $s$-wave neutron capture reaction were calculated as a 
function of relative  energy for both channel spins.
The results of these calculations are shown in Fig. 7.  The curve 
labeled (a) corresponds 
to the sum of the contribution of channel-spin $s=3/2$ and 
$s=5/2$,  where each contribution was calculated with different scattering 
potential.
The curves (b) correspond to a different assumption for the scattering
potential in which the well depths are the same as 
that for the bound state. The dotted curve is obtained considering
only an $s$-wave neutron, and the solid line is the sum of the 
$s$ and $d$-wave ontributions. As one can see in Fig.
7, the onset of the higher partial wave component is observable only 
at higher energies.

In order to compare our results for the 
$^8$Li(n,$\gamma$)$^9$Li$_{gs}$ capture reaction with other measurements 
and calculations, we have computed
the nucleosynthesis reaction rate. The expression  
for the reaction rate for E1 capture in cm$^3$mol$^{-1}$s$^{-1}$ is given 
by \cite{ang99}:\\

$N_A<\sigma v>=K\int_{0}^{\infty}\sigma(E)Eexp(-C_2E/T_9)dE$, 
\hspace{6mm} (3)\\

\noindent where \\

$K = C_1\mu^{-1/2}T_9^{-3/2}$
\\
\\
\noindent and $C_1=3.7313\times10^{10}$, $C_2=11.605$, 
N$_A$ is Avogadro's number, $\mu$ is the reduced mass of 
the system, $T_9$ is the temperature in units of 10$^9$~K, 
$\sigma$ is the capture cross section, $v$ is the relative velocity, and
$E$ is the energy in the center-of-mass system. $E$ is 
given in MeV and the cross section in $barns$. 

The reaction rate for the   $^8$Li(n,$\gamma$)$^9$Li$_{gs}$ capture reaction 
was estimated at a temperature $T_9$ = 1 and 
the integral of the expression [3] was performed up to 1.2 MeV.
At this temperature, the capture reaction becomes important
for the synthesis of heavier elements  in the inhomogeneous Big Bang model.
Also, the role of light neutron-rich nuclei for the r-process
in Type-II supernovae seems to be important at temperatures 
$0.5 <T_9<4$ \cite{ter01}.  Although the resonant 
capture through the ${5\over 2}^-$ resonance in $^9$Li could be important 
for temperatures higher than $T_9=0.5$,
in the present calculation only the direct 
capture to $^9$Li at the ground-state is considered. 

At energies up to 1 MeV, the capture reaction is completely dominated 
by $s$-wave neutrons. The contribution of the $d$-wave neutron 
to the reaction rate is estimated to be less than 0.5\% at these energies 
and it becomes important only at high energies. 
Assuming a distorted wave from
the potentials that have the same volume integral per nucleon as for the 
n+$^7$Li 
system, the reaction rate for 
the  $^8$Li(n,$\gamma$)$^9$Li$_{gs}$ capture reaction was deduced to be 
$N_A<\sigma v>=(3.17 \pm 0.70)\times 10^3~cm^3 mol^{-1}s^{-1}$, 
where the  uncertainty is from the uncertainty in the spectroscopic factor 
used in the calculation (20\%) and from the variation of $\pm$1 MeV in the 
potentials used to determined the distorted wave (10\%). 
As suggested by Mengoni, {\it et al.} \cite{men95}, a different assumption
would be to use the same 
potential for the incoming channel as that for the bound state. 
With this assumption for the n-$^8$Li potential, 
the reaction rate is determined to 
be $N_A<\sigma v>=(3.23 \pm 0.71)\times 10^3~cm^3 mol^{-1}s^{-1}$. 
The average value adopted here is:
\\ 
\\$N_A<\sigma v>=(3.20 \pm 0.70)\times 10^3~cm^3 mol^{-1}s^{-1}$.\\
\\ 
In Fig. 8, this value is compared with 
theoretical calculations 
reported in Refs. \cite{mal89,mao91,thi91,des93,rau94,bert99} and experimental 
estimations from Refs. \cite{zec98}, \cite{kob03} and \cite{li05}. 
Our result is comparable to the most recent theoretical calculations
\cite{thi91,des93,rau94,bert99} and is in good agreement with the
value from a recent (d,p) 
experiment \cite{li05}. 
The small difference obtained in the 
reaction rate between these two experiments can be attributed
in part to the different approaches used in the calculation of the
neutron scattering potential.  
Both results are significantly higher 
than the upper limit obtained in the most recent Coulomb dissociation 
experiment by Kobayashi,{\it et al.} \cite{kob03}.

A $1/v$ behavior of the capture cross section as 
a function of relative energy is expected for $s$-wave neutron capture
at low energies.
However, as can be seen in Fig. 7, the reaction rate obtained in 
the potential model follows a $1/v$ dependence only for very low 
energies ($E_n<0.1$ MeV). 
Normalizing a $1/v$ curve to the values 
given by the potential model in the energy range from 0-0.1 MeV,
the corresponding expression for the capture cross section becomes:
\\
\\$\sigma(n,\gamma)=3.95\times E_n^{-1/2}$,\\
\\
where $E_n$ is the neutron energy in MeV and $\sigma(n,\gamma)$ is given
in $\mu$b. 
The corresponding reaction rate integrated over the 
range up to 1.2 MeV (which would be in this case energy independent)
is:
\\
\\
$N_A<\sigma v>=3.27\times 10^3~cm^3 mol^{-1}s^{-1}$,
\\
\\
which is in agreement with the value obtained directly from the potential
model and also much higher than the upper limit obtained
by Kobayashi {\it et al.} \cite{kob03}.

\section{summary}

We have measured the angular distributions for the elastic scattering 
of $^8$Li on $^9$Be  and the neutron transfer reactions  
$^9$Be($^8$Li,$^7$Li)$^{10}$Be and $^9$Be($^8$Li,$^9$Li)$^8$Be at  
E$_{LAB}$ = 27.0 MeV. Spectroscopic factors for the 
$^8$Li$_{gs}\otimes$n=$^9$Li$_{gs}$ and 
$^7$Li$_{gs}\otimes$n=$^8$Li$_{gs}$ 
bound systems were obtained from the comparison 
between the experimental differential cross sections and FR-DWBA calculations
with the code FRESCO.   The spectroscopic factors obtained are compared with 
shell model calculations and also with experimental values from (d,p) 
reactions. 
Using the spectroscopic factors extracted from the angular distributions
for the $^8$Li$_{gs}\otimes$n=$^9$Li$_{gs}$ and  
$^7$Li$_{gs}\otimes$n=$^8$Li$_{gs}$ bound system, 
we have derived the cross-sections for the  $^7$Li(n,$\gamma$)$^8$Li and
$^8$Li(n,$\gamma$)$^9$Li$_{gs}$ neutron capture reactions based on a 
potential model. The reaction rates for the non-resonant
part of the $^8$Li(n,$\gamma$)$^9$Li$_{gs}$ reaction are compared with 
the results from previous 
indirect methods and with theoretical calculations. 
Our work has shown that low energy radioactive nuclear beams can be very 
suitable not only to perform spectroscopic investigations but also
to determine the non-resonant parts of capture reactions of 
astrophysics interest. 

\section*{Acknowledgments}

The authors wish to thank the Funda\c c\~ao de Amparo a Pesquisa
do Estado de S\~ao Paulo (FAPESP 2001/06676-9 and 2006/00629-2) 
for financial support.  This work was also funded in part by the
U.S. NSF under Grants No. PHY03-54828 and INT03-05347.

\newpage


\begin{table}
\caption{Optical-model potential parameters. 
Radii are given by $R_x~=~r_x \times A_T^{1/3}$.}
\begin{center}
\begin{tabular}{|ccccccccc|}
\hline
SET & $V^{a)}$  & $r_R$ & $a_R$ & $W_V^{a)}$  & $r_I$ & $a_I$ & $r_C$ &   \\
         & (MeV)  & (fm)  & (fm) & (MeV) & (fm)  & (fm)  & (fm) &   References \\
\hline
 1  & 173.1 & 1.19 & 0.78 & 8.90  & 2.52 & 0.924 & 1.78 & $^7$Li+$^9$Be at 34 
MeV \cite{kemp77}\\
 2  & 234.4 & 1.21 & 0.76 & 8.90  & 2.43 & 1.020 & 1.78 & $^7$Li+$^9$Be at 34 
MeV \cite{kemp77}\\
 3  & 152.0 & 1.38 & 0.75 & 6.72  & 2.72 & 0.900 & 1.20 & $^7$Li+$^9$Be at 24 
MeV \cite{perey76}\\
\hline
\end{tabular}
\end{center}
a) Volume-Wood-Saxon potential\\
\end{table}

\begin{table}
\caption{Spectroscopic factors $C^2S$.}
\begin{center}
\begin{tabular}{|lc|c|cc|c|}
\hline
                           &J$^{\pi}$& Shell Model       &                      &             & This 
work  \\  
                           &         & calculation       &  (d,p)               & (d,t)       & SET 1-2-3  
\\
\hline
$^8$Li$_{gs}\otimes$n=$^9$Li$_{gs}$    &  3/2-   & 0.628$^{a)}$ 0.885$^{b)}$ & 
0.68(14)$^{d)}$ 0.90$^{e)}$  &             & 0.62 (13)  \\
$^7$Li$_{gs}\otimes$n=$^8$Li$_{gs}$    &   2+    & 0.977$^{c)}$        &     
0.87$^{f)}$       &             & 1.03 (15)  \\
$^8$Be$_{gs}\otimes$n=$^9$Be$_{gs}$    &  3/2-   & 0.580$^{c)}$          &                     
&  0.44(7)$^{h)}$            &            \\
$^9$Be$_{gs}\otimes$n=$^{10}$Be$_{gs}$ &   0+    & 2.357$^{c)}$        &  
2.23 (13)$^{g)}$             &   &            \\
\hline
\end{tabular}
\end{center}
a) from Ref. \cite{nav04}\\
b) from Ref. \cite{jep06} using same Cohen Kurath wave-function  \\
c) from Cohen and Kurath \cite{coh67}\\
d) from (d,p) reaction at 39 MeV \cite{li05}\\
e) from (d,p) reaction at 76 MeV \cite{wuo05}\\
f) from Ref. \cite{nag05}\\
g) average of S=2.10 from Ref. \cite{dar76} and S=2.356 from Ref. \cite{pow70} 
\\
h) average of S=0.37 from Ref. \cite{dar76} and S=0.51 from Ref. \cite{fit67}\\
\end{table}

\begin{figure}
\epsfig{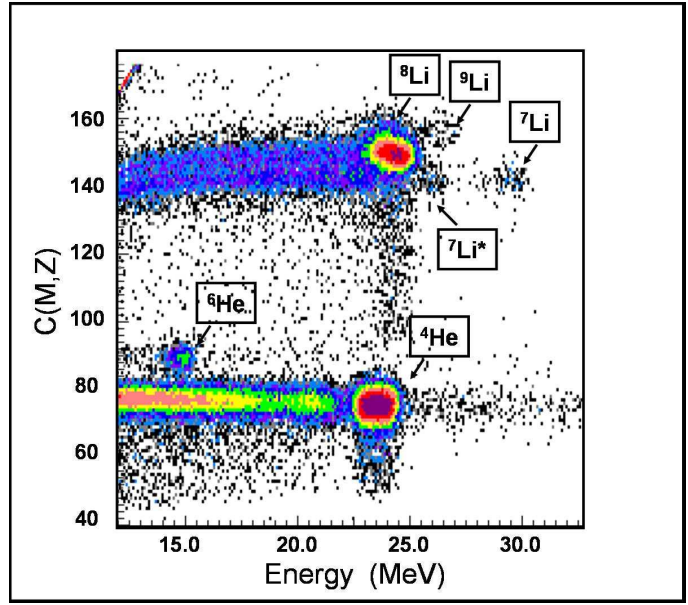}
\caption{Typical particle identification spectrum [$C(Z,M) \times E_{total}$]
showing the lithium and helium isotope region
for  the interaction of $^8$Li+$^9$Be at 15 degrees. The elastic scattering 
($^8$Li) and reaction products ($^7$Li and $^9$Li) 
are indicated, as well as the $^4$He and $^6$He contamination in the 
secondary beam.}
\end{figure}

\begin{figure}
\epsfig{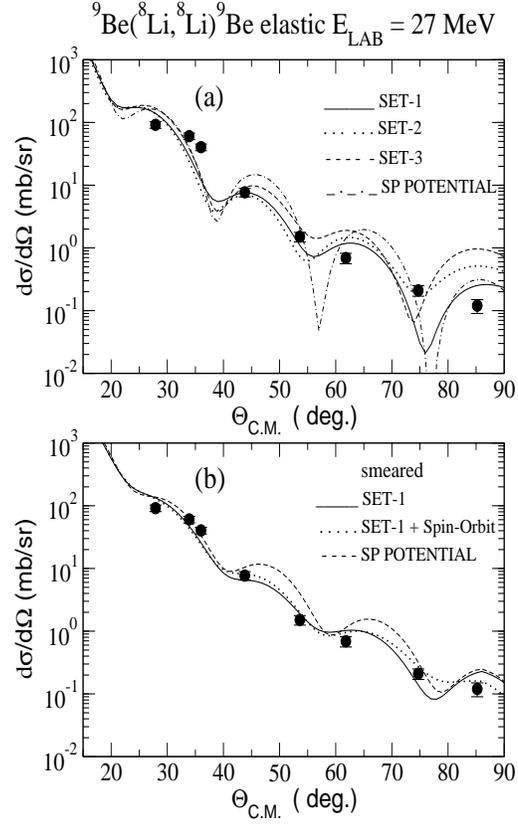}
\caption{The differential cross sections for elastic scattering 
($^9$Be($^8$Li,$^8$Li)$^{9}$Be) at 27 MeV incident $^8$Li lab energy. 
The curves are optical-model
calculations with the sets of parameters listed in Table I.}
\end{figure}

\begin{figure}
\epsfig{file=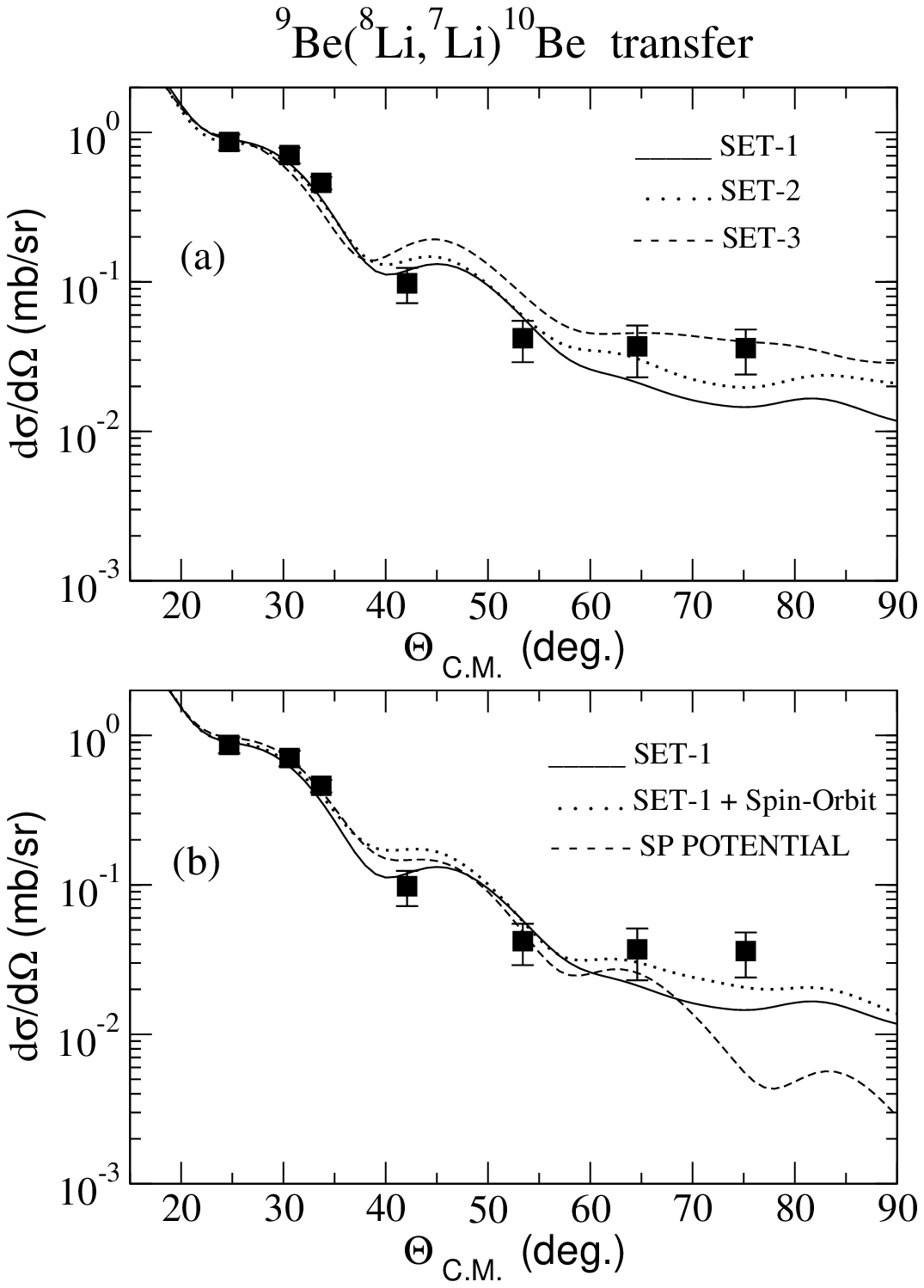,height=12cm,width=8cm}
\caption{The differential cross sections for the 
$^9$Be($^8$Li,$^7$Li$_{gs}$)$^{10}$Be$_{gs}$ neutron transfer reaction at 
27 MeV incident energy. The curves are FR-DWBA calculations using the code 
FRESCO with the potentials indicated (see Table I), as discussed in the text.}
\end{figure}

\begin{figure}
\epsfig{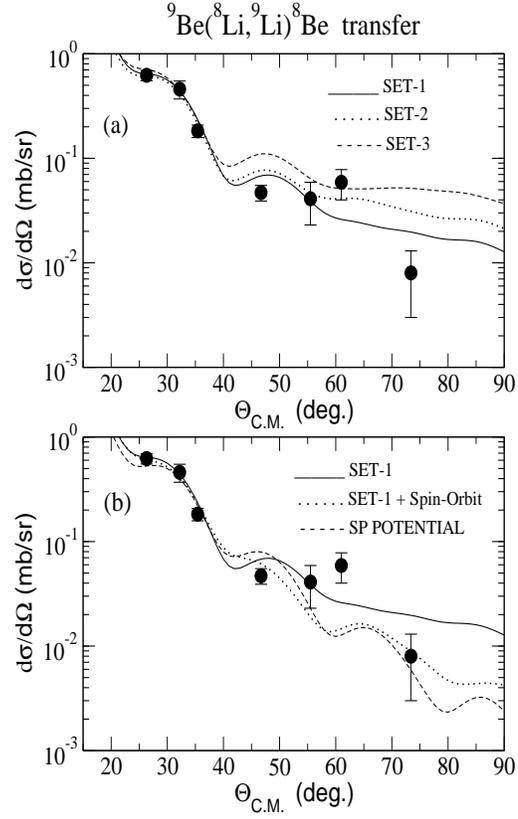}
\caption{The differential cross sections for the 
$^9$Be($^8$Li,$^9$Li$_{gs}$)$^{8}$Be$_{gs}$ neutron transfer reaction at 
27 MeV incident energy. The curves are FR-DWBA calculation using the code 
FRESCO with the potentials indicated, as discussed in the text.}
\end{figure}

\begin{figure}
\epsfig{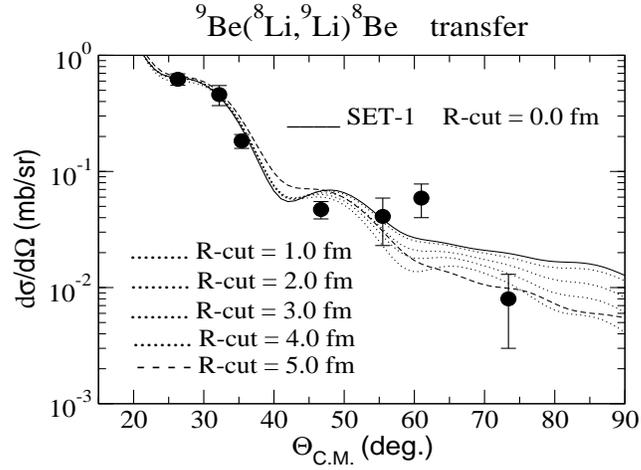}
\caption{Angular distribution for the 
$^9$Be($^8$Li,$^9$Li$_{gs}$)$^{8}$Be$_{gs}$ neutron transfer reaction at 
27 MeV incident energy.  The solid line corresponds to the FR-DWBA calculation 
with potential SET-1 (Table-I). The dotted lines are FR-DWBA calculations with
cutoffs in the radial form factor integral as indicated. 
The dashed curve is the FR-DWBA calculation with a cutoff radius of 5.0 fm}
\end{figure}

\begin{figure}
\epsfig{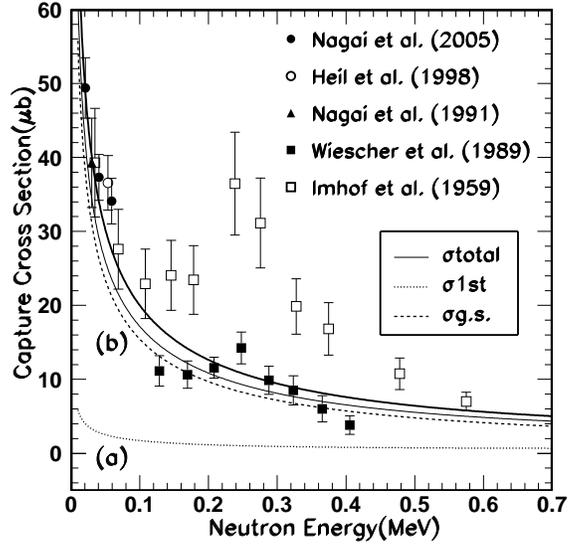}
\caption{The cross sections for the $^7$Li(n,$\gamma$)$^8$Li capture
reaction. The experimental points are from Refs. 
\cite{nag05} and \cite{hei98,nag91,wie89,imh59}. 
The curve labeled (a) is the sum of channel-spin $s=1$  and $s=2$ 
contributions for the neutron  capture reaction to the 
first excited state of $^8$Li, while curve (b) is the sum of the 
channel-spin $s=1$ and $s=2$ contributions for the $^8$Li ground-state.
The thin solid line is the sum of these two contributions 
using the spectroscopic factor $S_{^8Li}$(g.s.)=0.87. The thick solid line
is the same calculation but considering the spectroscopic
factor $S_{^8Li}$(g.s.)=1.03 (15) obtained from the present work.}
\end{figure}

\begin{figure}
\epsfig{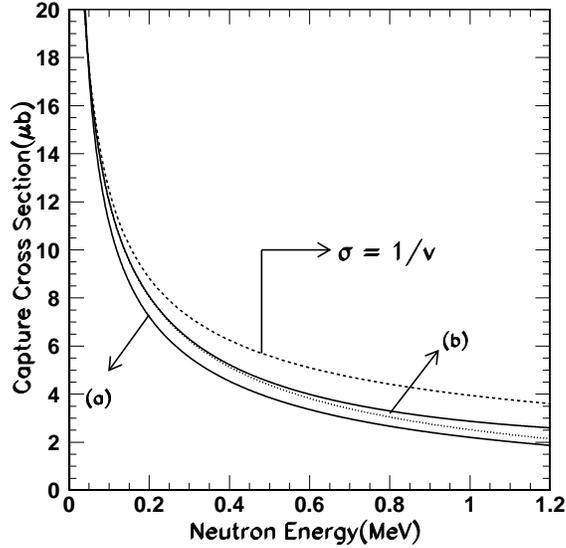}
\caption{The results of the capture cross section calculations
for the $^8$Li(n,$\gamma$)$^9$Li$_{gs}$ reaction.  The lower curves 
labeled (a) correspond  to the sum of channel-spin $s=3/2$ and 
$s=5/2$ assuming potential depths scaled from the 
n+$^7$Li capture reaction analysis, as explained in the text. 
Curves labeled (b) correspond to the assumption of same potential for the 
incoming wave-function as for the bound state, for $s$-wave
neutron only (dotted curve) and $s$ and $d$-wave neutrons (solid curve).
The uppermost curve (dashed line) corresponds to a $1/v$ fit
to the low-energy cross sections.} 
\end{figure}

\begin{figure}
\epsfig{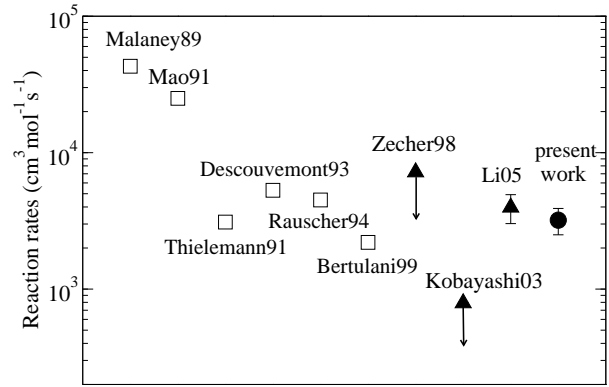}
\caption{Reaction rate for the direct neutron capture reaction
$^8$Li(n,$\gamma$)$^9$Li$_{gs}$ calculated at $T_9$=1. The open
squares are calculations from Refs.  
\cite{mal89,mao91,thi91,des93,rau94,bert99} while the full triangles are 
experimental  estimations from Refs. 
\cite{zec98}, \cite{kob03} and \cite{li05}.} 
\end{figure}

\end{document}